\documentclass[preprint,amsmath,amssymb,aps,nofootinbib,showpacs]{revtex4}
\usepackage{graphicx}
\usepackage{bm}
\usepackage{subfigure}
\usepackage{amsmath}

\newcommand{\epl}{Europhys. Lett.\ }

\newcommand{\pr}{Phys. Rev.\ }

\newcommand{\jpa}{J. Phys. A\ }
\newcommand{\jpb}{J. Phys. B\ }
\newcommand{\njp}{New J. Phys.\ }
\newcommand{\etal}{{\em et al. }}
\newcommand{\e}{\mbox{e}}

\newcommand{\UQ}{School of Mathematics and Physics, University of Queensland, Brisbane, 
QLD 4072, Australia.}

\begin{document}

\title{Quantum behaviour of open pumped and damped Bose-Hubbard trimers}

\author{C.~V. Chianca and M.~K. Olsen}
\affiliation{\UQ}

\date{\today}

\begin{abstract}

We propose and analyse analogs of optical cavities for atoms using three-well inline Bose-Hubbard models with pumping and losses.
With one well pumped and one damped, we find that both the mean-field dynamics and the quantum statistics show a qualitative dependence on the choice of damped well. The systems we analyse remain far from equilibrium, although most do enter a steady-state regime. We find quadrature squeezing, bipartite and tripartite inseparability and entanglement, and states exhibiting the EPR paradox, depending on the parameter regimes. We also discover situations where the mean-field solutions of our models are noticeably different from the quantum solutions for the mean fields. 
Due to recent experimental advances, it should be possible to demonstrate the effects we predict and investigate in this article.

\end{abstract}

\maketitle

\section{Introduction}
\label{sec:intro}

Recent experimental advances in the manipulation of ultracold atoms, particularly with respect to the configuration of almost arbitrary potentials~\cite{painting,Tylerpaint}, and the damping of individual wells of an optical lattice~\cite{NDC}, as well as the possibility of pumping a Bose-Hubbard system from a larger reservoir condensate~\cite{Kordas1,Kordas2}, have opened new possibilities in quantum atom optics. Along with theoretical schemes for the measurement of atomic quadratures~\cite{andyhomo,homoSimon}, these advances allow for the realisation of schemes which would not have been possible in the recent past. In particular, we can consider atomic equivalents of optical cavities, where the nonlinearity is inherent to the system. A related early proposal by Drummond and Walls analysed a quantum optical system consisting of a Kerr medium inside a Fabry-Perot cavity, which is mathematically the equivalent of a pumped and damped single isolated well of a Bose-Hubbard model~\cite{PDDDFW}, although Kerr nonlinearities tend to be lower with optical systems. This model was extended to a dimer by Olsen~\cite{nlc}, and to a trimer by Tan \etal~\cite{HTTan}, who both examined the quantum correlations in the output and intracavity fields. The atomic version of these systems has one qualitative difference from the optical versions in that not all wells must necessarily experience damping, which is not possible with optical cavities. This allows us to assign pumping and loss independently to the wells, using the techniques mentioned above.

In terms of atomic systems, Pi\u{z}orn has analysed Bose-Hubbard models with pumping and dissipation~\cite{Pizorn}, using techniques which are useful for moderate numbers of atoms and wells, while Cui \etal have investigated driven and dissipative Bose-Hubbard models, obtaining mean-field analytical results for a two-well system~\cite{Cui}. Kordas \etal have also analysed triangular trimers and inline chains with dissipation at one well~\cite{Kordas3,Kordas4}, finding some interesting physical effects. In work closely related to the present article, Olsen \etal~\cite{BHcav2} have analysed a two well Bose-Hubbard dimer with pumping and damping each at only one of the wells, and Olsen has looked at the same system in terms of non-Gaussian properties and Eintein-Podolsky-Rosen correlations~\cite{NGBHcav2}. Olsen and Bradley~\cite{BHAx} have analysed an open trimer system with pumping at one well and damping at the other two in terms of its performance as a quantum correlated twin atom laser.   

In this work we analyse both the dynamics and steady-state properties of inline Bose-Hubbard trimers~\cite{Nemoto,Chiancathermal} with pumping at the first well and damping at one of the three. We will use the truncated Wigner representation~\cite{Graham,Steel}, which does not impose a computational limitation on the number of atoms and has the advantage over matrix methods that the computational complexity scales linearly with the number of wells. Although the positive-P representation~\cite{P+} would be our preferred choice, and was used in the analysis of the trimer with damping at two of the wells~\cite{BHAx}, it suffers from catastrophic instabilities when only one of the three wells is damped. We note here that the truncated Wigner, while an approximation, performed well for the open dimer~\cite{NGBHcav2}, differing from the positive-P predictions in none of the observables that we will calculate here, and that we are not calculating two-time correlation functions, for which the truncated Wigner is known to be unreliable~\cite{turco2time}. We also note that we are dealing with an open system so that the quantum states will be mixed and the Wigner function will be strictly positive~\cite{Wigpos}. For these reasons, we fully expect the truncated Wigner representation to be accurate for this system.

\section{Physical model, Hamiltonian, and equations of motion}
\label{sec:model}

We consider three different geometric configurations of the open trimer, all with pumping at well one, and  the damping at well one, two, or three. 
For the generic system, the Bose-Hubbard~\cite{BHmodel,Jaksch,BHJoel} unitary Hamiltonian is
\begin{equation}
{\cal H} = \hbar\chi\sum_{i=1}^{3}\hat{a}_{i}^{\dag\,2}\hat{a}_{i}^{2}-\hbar J \left(\hat{a}_{1}^{\dag}\hat{a}_{2}+\hat{a}_{2}^{\dag}\hat{a}_{1} +\hat{a}_{2}^{\dag}\hat{a}_{3}+\hat{a}_{3}^{\dag}\hat{a}_{2} \right),
\label{eq:genHam3line}
\end{equation}
where $\hat{a}_{i}$ is the bosonic annihilation operator for the $i$th well, $\chi$ represents the collisional nonlinearity and $J$ is the tunneling strength.
We will always consider that the pumping is of well $1$ and can be represented by the Hamiltonian
\begin{equation}
{\cal H}_{pump} = i\hbar\left(\hat{\Gamma}\hat{a}_{1}^{\dag}-\hat{\Gamma}^{\dag}\hat{a}_{1}\right),
\label{eq:pump}
\end{equation}
which is commonly used for the investigation of optical cavities. The basic assumption here is that the first well receives atoms from a coherent condensate which is much larger than any of the modes in the wells we are investigating, so that it will not become depleted over the time scales of interest. It is important to state that when we refer to steady-state values, we are referring to stationary solutions over the appropriate time scale.
The damping term for well $i$ acts on the system density matrix via the Lindblad superoperator
\begin{equation}
{\cal L}\rho = \gamma\left(2\hat{a}_{i}\rho\hat{a}_{i}^{\dag}-\hat{a}_{i}^{\dag}\hat{a}_{i}\rho-\rho\hat{a}_{i}^{\dag}\hat{a}_{i}\right),
\label{eq:damp}
\end{equation}
where $\gamma$ is the coupling between the damped well and the atomic bath, which we assume to be unpopulated.
If the lost atoms fall under gravity, we are justified in using the Markov and Born approximations~\cite{JHMarkov}.

Following the usual procedures~\cite{QNoise,DFW}, we may map the von Neumann equation and the master equation onto a generalised Fokker-Planck equation in the Wigner representation. This is not a true Fokker-Planck equation because it has third-order derivatives and, although it can be mapped onto stochastic difference equations~\cite{nossoEPL}, the numerical integration of these is extremely unstable. By dropping the third-order terms, usually under the assumption that they are small, we may map the problem onto stochastic equations. For this system, the Stratonovich and It\^o stochastic equations~\cite{SMCrispin} have the same form since the noise is additive. As an example, the equations for an open trimer with pumping at well $1$ and loss at well $3$ are
\begin{eqnarray}
\frac{d\alpha_{1}}{dt} &=& \epsilon - 2i\chi |\alpha_{1}|^{2}\alpha_{1}+iJ\alpha_{2}, \nonumber \\
\frac{d\alpha_{2}}{dt} &=& -2i\chi|\alpha_{2}|^{2}\alpha_{2}+iJ(\alpha_{1}+\alpha_{3}), \nonumber \\
\frac{d\alpha_{3}}{dt} &=& -\gamma\alpha_{3}-2i\chi|\alpha_{3}|^{2}\alpha_{3}+iJ\alpha_{2} +\sqrt{\gamma}\eta ,
\label{eq:cav3line}
\end{eqnarray}
where $\epsilon$ represents the rate at which atoms enter well $1$, $\gamma$ is the loss rate from the selected well, and $\eta$ is a complex Gaussian noise with the moments $\overline{\eta(t)}=0$ and $\overline{\eta^{\ast}(t)\eta(t')}=\delta(t-t')$. The variables $\alpha_{i}$ correspond to the operators $\hat{a}_{i}$ in the sense that averages of products of the Wigner variables over many stochastic trajectories become equivalent to symmetrically ordered operator expectation values, for example $\overline{|\alpha_{i}|^{2}}=\frac{1}{2}\langle\hat{a}_{i}^{\dag}\hat{a}_{i}+\hat{a}_{i}\hat{a}_{i}^{\dag}\rangle$. The initial states in all wells are vacuum, sampled as in Olsen and Bradley~\cite{states} for coherent states with vacuum excitation. We will use $\epsilon=10$ and $\gamma=J=1$ in all our numerical investigations, while using two values of $\chi$, $10^{-3}$ and $10^{-2}$. The equations for configurations with damping at a different well are found by the simple transfer of the terms involving $\gamma$. The truncated Wigner equations of motion are then integrated numerically to obtain close approximations to expectation values of the desired operator moments. In this work, at least $10^{5}$ trajectories were averaged over, which gave good convergence with minimal sampling error.

\section{Quantities of interest}
\label{sec:interest}      

There are several quantities of interest here, including the populations in each well, $\overline{|\alpha_{i}|^{2}}-\frac{1}{2}$, the quadrature variances, various entanglement correlations, and the possibility of Einstein-Podolsky Rosen (EPR) steering. We will now give the definitions for all the quantities reported on in what follows.
We start by giving the quadrature definition we will use, since this effects the values of the various inequalities used for squeezing, inseparability, entanglement and EPR tests. We define a general quadrature as
\begin{eqnarray}
\hat{X}_{j}(\theta) &= & \hat{a}_{j}\e^{-i\theta}+\hat{a}_{j}^{\dag}\e^{i\theta},
\label{eqn:Xtheta}
\end{eqnarray}
so that the $\hat{Y}_{j}(\theta)=\hat{X}_{j}(\theta+\pi/2)$, squeezing exists whenever a quadrature variance is found to be less than $1$, for any angle. As is well known, one of the effects of a $\chi^{(3)}$ nonlinearity is to cause any squeezing to be found at a non-zero quadrature angle~\cite{nlc}. 

Having defined our quadratures, we may now define the correlations we will investigate to detect bipartite mode inseparability. The first of these, known as the Duan-Simon inequality~\cite{Duan,Simon}, states that, for any two separable states,
\begin{equation}
V(\hat{X}_{j}+\hat{X}_{k})+V(\hat{Y}_{j}-\hat{Y}_{k}) \geq 4,
\label{eq:DS}
\end{equation}
with any violation of this inequality demonstrating the inseparability of modes $j$ and $k$. For the mixed states we consider here, this violation does not necessarily prove entanglement, but the next correlation considered does. 

This correlation is that of the Einstein-Podolsky-Rosen (EPR) paradox~\cite{Einstein}, now also known as EPR-steering~\cite{Erwin,Wiseman}, and often detected by what has become known as the Reid criterion~\cite{EPRMDR}. Firstly we define the inferred quadrature variances of  two bosonic modes labelled $i$ and $j$, with an observer of mode $j$ inferring values of mode $i$, as 
\begin{eqnarray}
V_{inf}(\hat{X}_{i}) = V(\hat{X}_{i})-\frac{\left[V(\hat{X}_{i},\hat{X}_{j})\right]^{2}}{V(\hat{X}_{j})}, \nonumber \\
V_{inf}(\hat{Y}_{i}) = V(\hat{Y}_{i})-\frac{\left[V(\hat{Y}_{i},\hat{Y}_{j})\right]^{2}}{V(\hat{Y}_{j})},
\label{eq:VXYinf}
\end{eqnarray}
where we have suppressed the $\theta$ dependence for ease of notation and $V(AB) = \langle AB\rangle-\langle A\rangle\langle B\rangle$. If these were genuine physical variances, the Heisenberg Uncertainty Principle would require that
\begin{equation}
V_{inf}(\hat{X}_{i})V_{inf}(\hat{Y}_{i}) \geq 1.
\label{eq:HUPXY}
\end{equation}
As shown by Reid, a violation of this inequality signifies a two-mode state which demonstrates the EPR paradox, which necessarily means that the two modes are entangled since EPR states are a subset of the entangled ones. It was subsequently noticed that this criterion is directional~\cite{Wiseman}, with the ability to swap $i$ and $j$ in Eq.~\ref{eq:VXYinf}. This raised the possibility that the considered state could be asymmetric, with the two observers not agreeing on whether the bipartite system exhibited EPR-steering or not. This asymmetric steering has been predicted in optical~\cite{SFG,Sarah,asymSHG} and atomic systems~\cite{TWTEPR}, and measured in the laboratory~\cite{Handchen}, all using Gaussian measurements. It is now established that it is a general property, and may exist for any possible measurements~\cite{oneway}. In what follows, we will denote the value of the product of the inferred variances as $EPR_{ij}$ when the quadrature variances of mode $i$ are inferred by measurements at mode $j$.

For three mode inseparability, we may use the van-Loock Furusawa inequalities~\cite{vLF} 
\begin{equation}
V_{ij} = V(\hat{X}_{i}-\hat{X}_{j})+V(\hat{Y}_{i}+\hat{Y}_{j}+g_{k}\hat{Y}_{k}) \geq 4, 
\label{eq:VLF}
\end{equation}
for which the violation of any two demonstrates tripartite inseparability. The $g_{j}$, which are arbitrary and real, can be optimised~\cite{AxMuzz}, using the variances and covariances, as
\begin{equation}
g_{i} = -\frac{V(\hat{Y}_{i},\hat{Y}_{j})+V(\hat{Y}_{i},\hat{Y}_{k})}{V(\hat{Y}_{i})},
\label{eq:VLFopt}
\end{equation}
which is the process we follow here. Teh and Reid~\cite{Teh&Reid} subsequently showed that, for mixed states, tripartite entanglement is demonstrated if the sum of the three correlations is less than $8$, with genuine tripartite EPR-steering requiring a sum of less than $4$. 

Another set of inequalities was also presented by van Loock and Furusawa, the violation of any one of which is sufficient to prove tripartite inseparability,
\begin{equation}
V_{ijk} = V(\hat{X}_{i}-\frac{\hat{X}_{j}+\hat{X}_{k}}{\sqrt{2}})+V(\hat{Y}_{i}+\frac{\hat{Y}_{j}+\hat{Y}_{k}}{\sqrt{2}}) \geq 4.
\label{eq:VLFijk}
\end{equation}
Following the work of Teh and Reid again, for mixed states any one of these less than $2$ demonstrates genuine tripartite entanglement, while one of them less than $1$ demonstrates genuine tripartite EPR steering. How these inequalities work in practice for a given system has been demonstrated by Olsen and Corney~\cite{OCbitri}, for both pure and mixed states. The extra features available in a tripartite system where not all bipartitions are equal has also been used to show how an asymmetric system could be used to enable different levels of security inside one quantum key distribution network~\cite{promiscuity}.

The possibility of tripartite states which exhibit EPR-steering has also been investigated, with Wang \etal showing that the steering of a given quantum mode is allowed when not less than half of the modes within the states take part in the steering group~\cite{halfplus}. In a tripartite system, this means that measurements on two of the modes are needed to steer the third. In order to quantify this, we will use correlation functions developed by Olsen, Bradley, and Reid~\cite{OBR}, again using inferred quadrature variances. With the tripartite inferred variances as
\begin{eqnarray}
V_{inf}^{(t)}(\hat{X}_{i}) &=& V(\hat{X}_{i})-\frac{\left[V(\hat{X}_{i},\hat{X}_{j}\pm\hat{X}_{k})\right]^{2}}{V(\hat{X}_{j}\pm\hat{X}_{k})}, \nonumber \\
V_{inf}^{(t)}(\hat{Y}_{i}) &=& V(\hat{Y}_{i})-\frac{\left[V(\hat{Y}_{i},\hat{Y}_{j}\pm\hat{Y}_{k})\right]^{2}}{V(\hat{Y}_{j}\pm\hat{Y}_{k})}, 
\label{eq:V3inf}
\end{eqnarray}
we define the correlation function
\begin{equation}
OBR_{ijk} = V_{inf}^{(t)}(\hat{X}_{i})V_{inf}^{(t)}(\hat{Y}_{i}),
\label{eq:OBR}
\end{equation}
so that a value of less than one means that mode $i$ can be steered by the combined forces of modes $j$ and $k$.
On a final note, we mention that all the quantities needed for the correlations above can in principle be measured, either by density (number) measurements or via atomic homodyning~\cite{andyhomo,homoSimon}. 
We will now proceed to present results for the quantities of interest for each of the three configurations.

\section{Damping at third well}
\label{sec:g3line}

\begin{figure}[tbhp]
\centering
\begin{minipage}{.5\textwidth}
  \centering
  \includegraphics[width=.9\linewidth,height=0.65\linewidth]{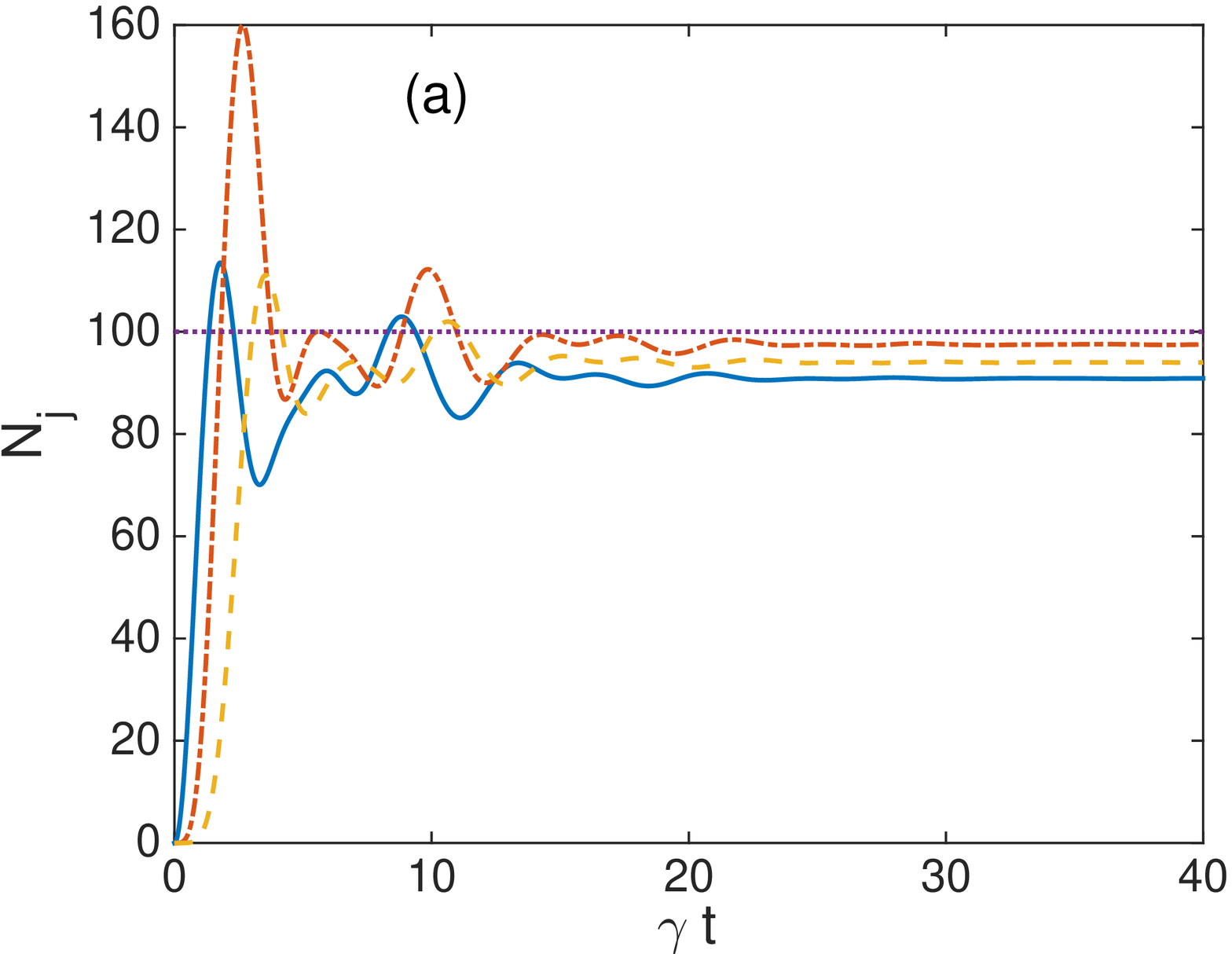}
\end{minipage}%
\begin{minipage}{.5\textwidth}
  \centering
  \includegraphics[width=.9\linewidth,height=0.65\linewidth]{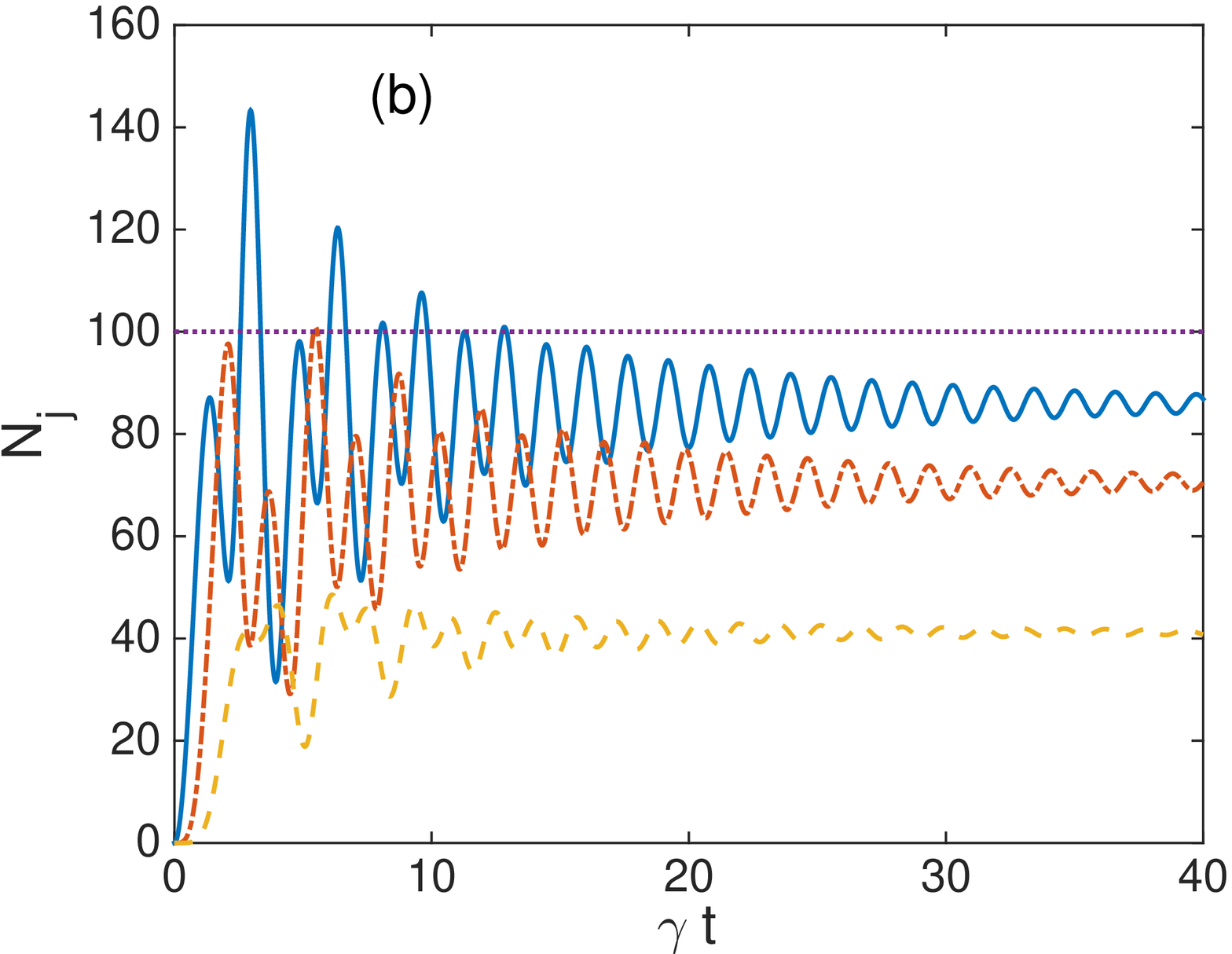}
 \end{minipage}
 \caption{(Color online)(a) The populations in each well as a function of time for $\chi=10^{-3}$ and damping at well 3. The solid line is $N_{1}$, the dash-dotted line is $N_{2}$, and the dashed line is $N_{3}$. The dotted line represents the non-interacting steady-state analytical solutions, all of which are equal for this configuration. The quantities plotted in this and subsequent plots are dimensionless. \newline
 (b) The populations in each well as a function of time for $\chi=10^{-2}$ and damping at well 3. The line styles are the same as in (a).}
 \label{fig:popsg3line}
\end{figure}

If we set $\chi=0$, we can find the classical steady-state analytical solutions
\begin{eqnarray}
\alpha_{1} &=& \frac{\epsilon}{\gamma}, \nonumber \\
\alpha_{2} &=& \frac{i\epsilon}{J}, \nonumber \\
\alpha_{3} &=& -\frac{\epsilon}{\gamma}.
\label{eq:classicp1g3}
\end{eqnarray}
Although we are not able to find exact analytical solutions for $\chi\neq 0$, the solutions above do give some insight. For example, they show that the phase in each well is different, but that the intensities, or atom numbers, in the steady state are all equal. As we can see from Fig.~\ref{fig:popsg3line}, when a finite nonlinearity is included, this no longer holds. Fig.~\ref{fig:popsg3line}(a) shows that for the lower of the two nonlinearities considered,  the well occupations settle down after some initial transient behaviour to steady-state values which are not equal, and are a little less than the non-interacting values. Fig.~\ref{fig:popsg3line}(b) shows that, for the higher nonlinearity of $\chi=10^{-2}$, the populations are even more suppressed from the non-interacting value, and that oscillations continue over a longer time, although they are damped. Note that we have plotted all populations up to a time of $\gamma t=40$, so that the initial transient behaviour can be clearly seen and that the oscillations seen in the second graph do damp out to a steady-state at a longer time. The tendency for populations to decrease with increasing interaction strength can be explained when we recognise that the interaction causes the phase rotation typical of Kerr nonlinearites so that the phase of each mode is longer as in Eq.~\ref{eq:classicp1g3}. This can act in a similar way to either detuning in an optical cavity or the Kerr nonlinearities in a nonlinear coupler~\cite{nlc}, and thus deplete the populations. In the nonlinear coupler with both wells equally pumped and damped, both modes will have the same steady-state phase, but this is not the case here since each well is in a different environment, some being either unpumped or undamped. This is a degree of freedom not available with coupled optical cavities, where all modes are necessarily damped. The differences in the phases of the mode in each well will be seen to have an effect when we calculate the optimal phase angles for the quantum correlations that we will analyse below.  

The table below shows the single mode quadrature variances and bipartite Duan-Simon correlations for this configuration, for both nonlinearities considered, and at the quadrature angles of their minimum values. Note that, for the higher nonlinearity, these values were computed at the longer time of $\gamma t=80$, after the oscillations seen in Fig.~\ref{fig:popsg3line}(b) had settled down. We see that for the lower nonlinearity, all these values are in the classical regime, with neither quadrature squeezing nor bipartite entanglement being found. When the nonlinearity is increased, we find that all three modes exhibit quadrature squeezing, while modes one and two now violate the Duan-Simon inequality. This difference for the two nonlinearities can be explained qualitatively by considering the effects of Kerr nonlinearities on the time evolution of quantum statistics~\cite{NGJoel}, in combination with their effect on tunneling. Because both systems are pumped at the same rate and the steady-state populations of each well are lower for the higher nonlinearity, the individual atoms may spend less time in each well on average than for the lower nonlinearity. The time they do spend is better in terms of the parameter $N_{j}\chi t$ for the optimisation of squeezing and entanglement. There would obviously be an optimal relationship between pumping, damping and nonlinearity for the best quantum correlations, but without analytical solutions this would be difficult to find.

\begin{center}
\begin{tabular}{ | c || c | c | c | c | c | c |}
\hline
 \multicolumn{7}{| c |}{Bipartite entanglement, loss at 3} \\
 \hline
   &  $V(\hat{X}_{1})$ & $V(\hat{X}_{2})$ & $V(\hat{X}_{3})$ & $DS_{12}$ & $DS_{13}$ & $DS_{23}$  \\ 
 \hline
 \hline
$\chi = 10^{-3}$ & 5.9@130$^o$ & 1.4@89$^o$  & 4.8@53$^o$ & 16.1@122$^o$ & 40.1@98$^o$ & 14.0@93$^o$ \\
\hline
 $\chi = 10^{-2}$ & 0.6@116$^o$   & 0.6@151$^o$   & 0.7@31$^o$   & 3.1@135$^o$  & 5.7@93$^o$  & 4.@176$^o$ \\
 \hline
 \hline
\end{tabular}
\end{center}

\begin{figure}[tbhp]
\includegraphics[width=0.75\columnwidth]{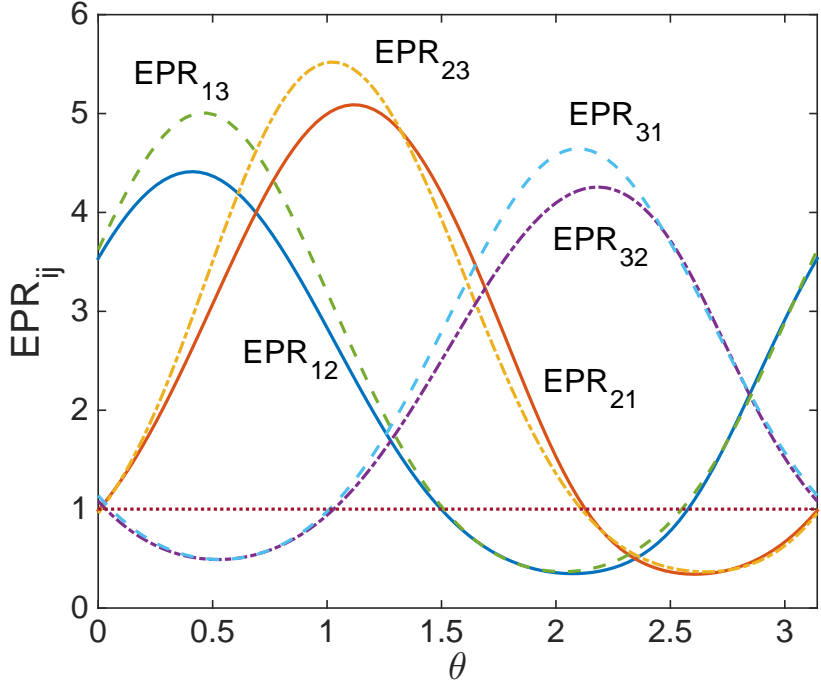}
\caption{(colour online) The steady-state Reid EPR correlations as a function of quadrature angle, for damping at well 3 and $\chi=10^{-2}$. The solid lines are for the bipartition of modes 1 and 2, from which it is seen that $EPR_{12}$ and $EPR_{21}$ can indicate EPR-steering simultaneously over a range of $\theta$. The dash-dotted lines are for modes 2 and 3, and neither these nor the dashed lines representing the bipartition of 1 and 3 are below the dotted line together at any quadrature angle. These two bipartitions exhibit an asymmetric steering which can be tuned by the angle chosen for measurement.}
\label{fig:EPRg3ki2}
\end{figure}

In the next table we show the bipartite EPR-steering results for this system, using the Reid criteria~\cite{EPRMDR} of Eq.~\ref{eq:VXYinf}. For the lower nonlinearity, we find no evidence of EPR-steering, whereas for $\chi=10^{-2}$ we find that all of the three possible bipartitions satisfy the criterion, but that the optimal angle for $EPR_{ij}$ is different than that for $EPR_{ji}$ in each case. We investigate this angular dependence further in Fig.~\ref{fig:EPRg3ki2}, where we see that only $EPR_{12}$ and $EPR_{21}$ violate the inequality over the same angles. The other two bipartitions never show EPR-steering for the same quadrature angles, which is a type of asymmetric steering~\cite{SFG,Sarah,asymSHG,TWTEPR}, although we should note that this is for Gaussian measurements and is of a type which, as far as we are aware, has not been considered previously.  This asymmetry due to quadrature angle is a feature which could be useful in some applications.

\begin{center}
\begin{tabular}{ | c || c | c | c | c | c | c |}
\hline
 \multicolumn{7}{| c |}{Bipartite EPR-steering, loss at 3} \\
 \hline
   &  $EPR_{12}$ & $EPR_{21}$ & $EPR_{23}$ & $EPR_{32}$ & $EPR_{13}$ & $EPR_{31}$  \\ 
 \hline
 \hline
$\chi = 10^{-3}$ & 33.2@127$^o$   & 1.4@38$^o$   & 1.8@88$^o$   & 18.5@58$^o$  & 6.6@124$^o$  & 4.9@58$^o$ \\
\hline
 $\chi = 10^{-2}$ & 0.4@118$^o$   & 0.4@150$^o$   & 0.4@153$^o$  & 0.5@29$^o$ & 0.5@31$^o$  & 0.4@116$^o$ \\
 \hline
 \hline
\end{tabular}
\end{center}

In the next table we consider the criteria for tripartite inseparability. We first note that none of the sum criteria of Teh and Reid~\cite{Teh&Reid} for the $V_{ij}$ can be used to show that the system exhibits either genuine tripartite entanglement or EPR steering for either nonlinearity. For the lower collisional interaction, neither of the tripartite inequalities are violated at all, while for $\chi=10^{-2}$ we see that one of the $V_{ij}$ is less than four. However, since two of these must be less than four for tripartite inseparabilty, this by itself is not sufficient. $V_{231}$ is slightly less than $4$, at $3.9$, so that tripartite inseparability is found by this measure. These measures do not demonstrate that either genuine tripartite entanglement or tripartite EPR-steering exist for this configuration, since these respectively require one of the $V_{ijk}$ to be less than either $2$ or $1$.

\begin{center}
\begin{tabular}{ | c || c | c | c | c | c | c |}
\hline
 \multicolumn{7}{| c |}{Tripartite entanglement, loss at 3} \\
 \hline
   &  $V_{12}$ & $V_{13}$ & $V_{23}$ & $V_{123}$ & $V_{231}$ & $V_{312}$   \\ 
 \hline
 \hline
$\chi = 10^{-3}$ & 14.4@118$^o$ & 39.9@98$^o$ & 13.2@62$^o$ & 31@102$^o$  & 8.4@96$^o$   & 33@87$^o$  \\
\hline
 $\chi = 10^{-2}$ & 2.9@135$^o$  & 5.2@82$^o$   & 4.2@176$^o$  & 4.3@145$^o$  & 3.9@151$^o$ & 4.9@153$^o$ \\
 \hline
 \hline
\end{tabular}
\end{center}

In this case we find that the tripartite EPR-steering inequalities of Eq.~\ref{eq:OBR} have more success. In the next table we show that, for $\chi=10^{-3}$, modes $3$ and $1$ can be used to steer mode $2$, while for $\chi=10^{-2}$, any of the bipartitions can steer the remaining mode. In this case, genuine EPR-steering does exist, although it is not detected by the other two methods. The success of some inequalities in a given situation where others may be unsuccessful in detecting either genuine tripartite entanglement or EPR-steering is always possible in a mixed state non-Gaussian system, and a similar phenomenon for bipartite correlations was found in a two-well pumped and damped Bose-Hubbard model~\cite{NGBHcav2,BHcav2}, where the Reid inequalities were violated in parameter regimes where the Duan-Simon inequalities were not.

\begin{center}
\begin{tabular}{ | c || c | c | c |}
\hline
 \multicolumn{4}{| c |}{Tripartite EPR-steering, loss at 3} \\
 \hline
   &   $OBR_{123}$ & $OBR_{231}$ & $OBR_{312}$  \\ 
 \hline
 \hline
$\chi = 10^{-3}$ &10.3@124$^o$  & 1.8@84$^o$  & 6.0@60$^o$    \\
\hline
 $\chi = 10^{-2}$ & 0.39@118$^o$  & 0.35@153$^o$ & 0.50@31$^o$\\
 \hline
 \hline
\end{tabular}
\end{center}

 \section{Damping at the middle well}
\label{sec:BH3b1g2}

\begin{figure}[tbhp]
\centering
\begin{minipage}{.5\textwidth}
  \centering
  \includegraphics[width=.9\linewidth,height=0.65\linewidth]{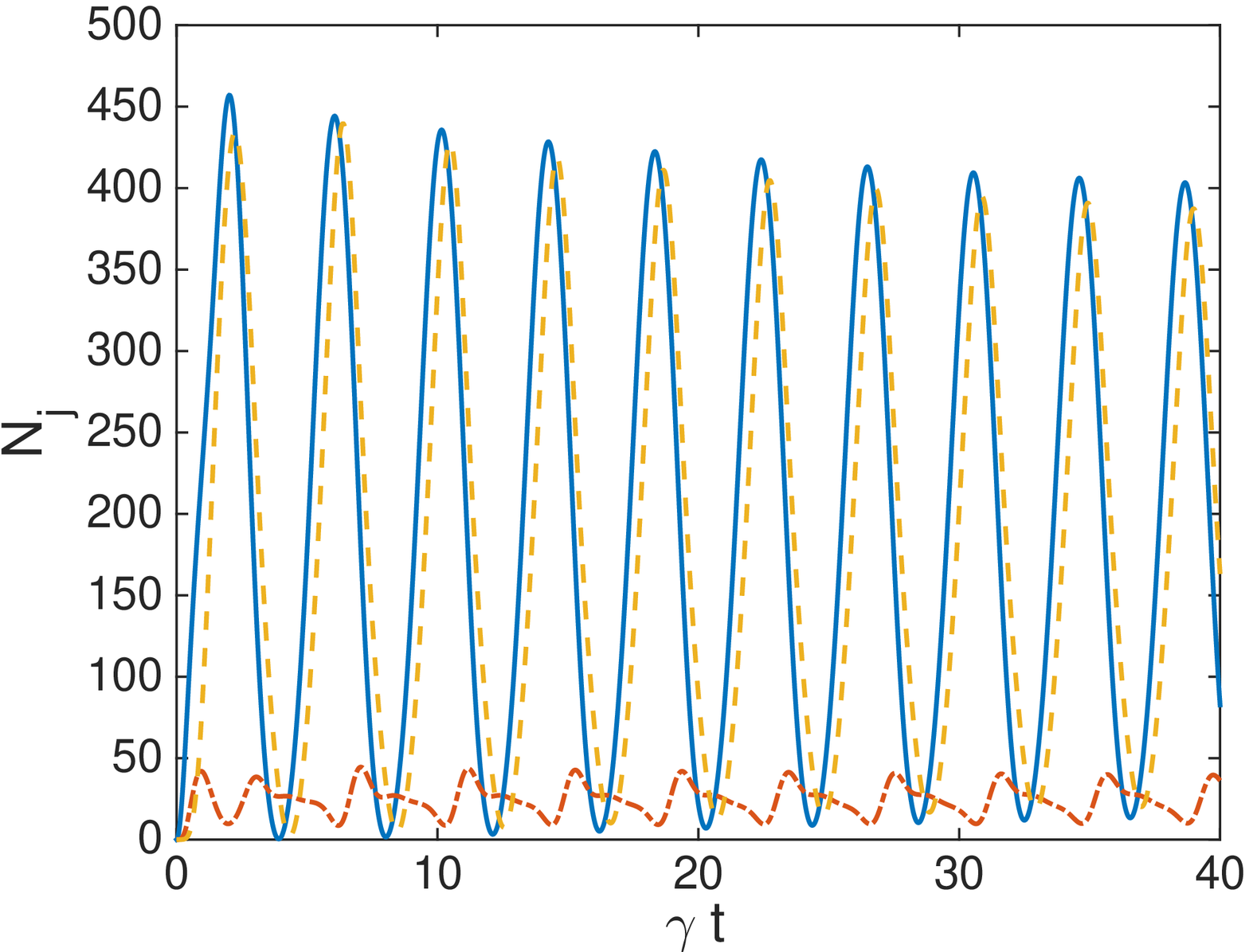}
\end{minipage}%
\begin{minipage}{.5\textwidth}
  \centering
  \includegraphics[width=.9\linewidth,height=0.65\linewidth]{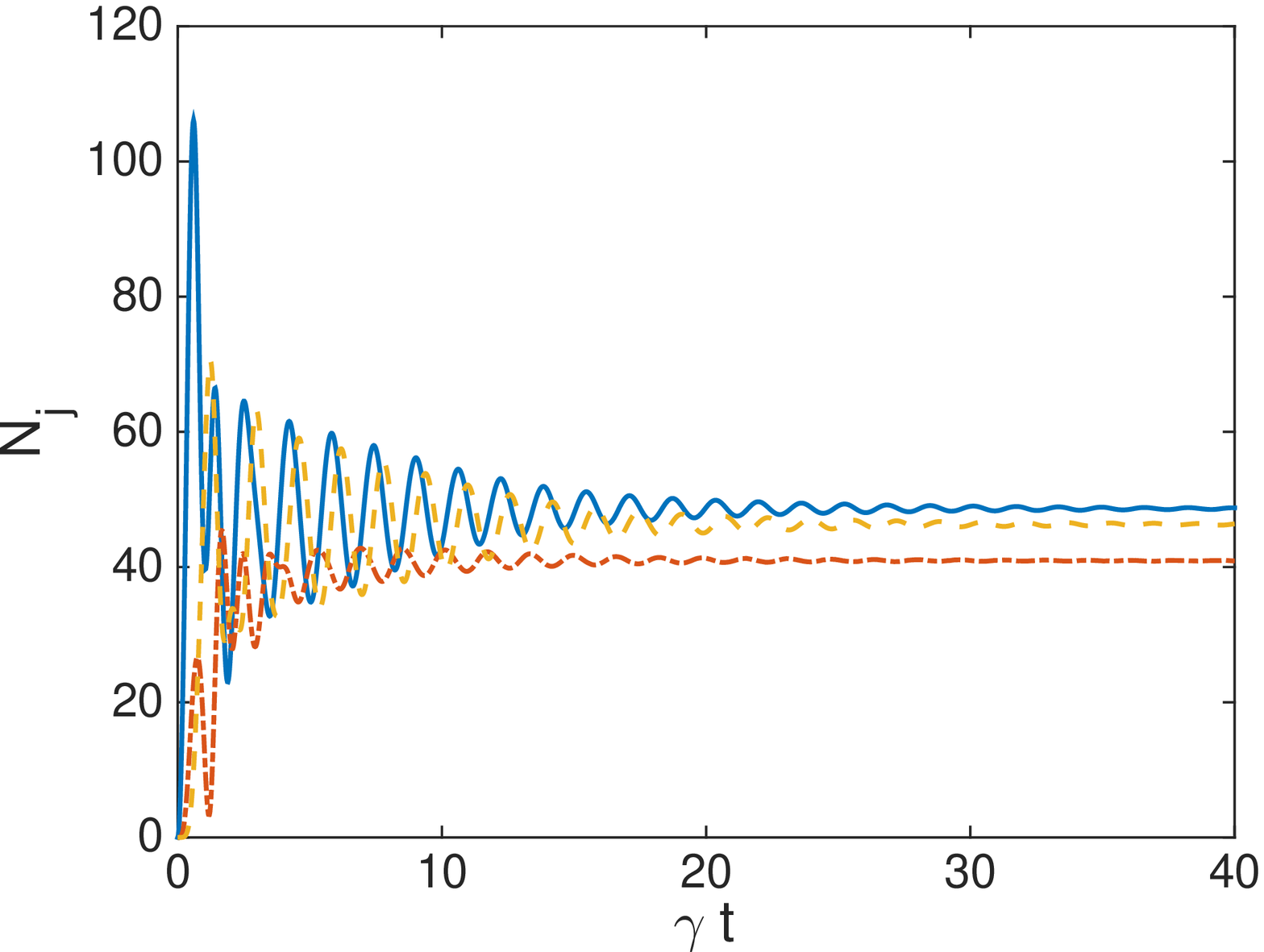}
 \end{minipage}
 \caption{(Color online)(a) The populations in each well as a function of time for $\chi=10^{-3}$ and damping at well 2. The solid line is $N_{1}$, the dash-dotted line is $N_{2}$, and the dashed line is $N_{3}$. \newline
 (b) The populations in each well as a function of time for $\chi=10^{-2}$ and damping at well 2. The line styles are the same as in (a).}
 \label{fig:popsg2line}
\end{figure}

In this case we find that the mean field dynamics have undergone a quantitative change and there are no non-interacting steady-state classical solutions except for $N_{2}$. This can be explained  when we write the non-interacting classical equations of motion im matrix form for the vector $\alpha=[\alpha_{1},\alpha_{2},\alpha_{3}]^{T}$,
\begin{equation}
\frac{d}{dt}\alpha = A\alpha+\epsilon,
\label{eq:Eqmat}
\end{equation}
where $\epsilon = [\epsilon,0,0]^{T}$ and
\begin{equation}
A =
\begin{bmatrix}
 0 & iJ & 0 \\
iJ &-\gamma & iJ \\
0 & iJ & 0
 \end{bmatrix}.
\label{eq:BH3g2mat}
\end{equation}
We see that $A$ has two identical rows and therefore is not invertible. When we solve for the classical non-interacting equations numerically, we find that the populations in one and three increase quadratically, while $N_{2}$ grows from zero then remains at approximately $25$. A finite $\chi$ stabilises this to something periodic in the classical case, but for small $\chi$ the maximum populations are well beyond the validity of the Bose-Hubbard model. In the classical case we don't find a steady-state until $\chi\approx 10^{-1}$ and this strength of interaction is beyond the scope of the present work. When we consider the truncated Wigner equation solutions, we find that there is no steady-state for $\chi=10^{-3}$, as shown in Fig.~\ref{fig:popsg2line}(a). For this reason, we cannot give steady-state quantum correlations for this value of the collisional interaction. However, when $\chi$ is increased to $10^{-2}$, we do find a steady-state solution after some initial transients, as shown in Fig.~\ref{fig:popsg2line}(b). The disagreement between the quantum and classical mean-field solutions here is even more striking than was found in the case of two Bose-Hubbard dimers coupled to each other via tunneling rates less than those within each dimer~\cite{BH4}. The damping of the oscillations for the increased collisional interaction can be explained in terms of the dephasing of different number state components, akin to that responsible for collapses and revivals in the Kerr oscillator~\cite{Agarwal}.  

The following table shows the steady-state quadrature variances and bipartite inseparability correlations for this configuration with $\chi=10^{-2}$. It shows that the modes are totally separable and that there is no quadrature squeezing, despite the fact that the quantum mean-field solutions are totally different to the classical ones, which keep oscillating. While the presence of a steady-state solution is clearly a quantum feature, this quantumness is not reflected in the bipartite correlations, which reflect uncorrelated noise in the system. 

\begin{center}
\begin{tabular}{ | c || c | c | c | c | c | c |}
\hline
 \multicolumn{7}{| c |}{Bipartite entanglement, loss at 2} \\
 \hline
   &  $V(\hat{X}_{1})$ & $V(\hat{X}_{2})$ & $V(\hat{X}_{3})$ & $DS_{12}$ & $DS_{13}$ & $DS_{23}$  \\ 
 \hline
\hline
 $\chi = 10^{-2}$ & 5.8@127$^o$  & 1.4@87$^o$   & 4.8@53$^o$    & 16.6@122$^o$  & 40.2@98$^o$  & 14.0@58$^o$  \\
 \hline
 \hline
\end{tabular}
\end{center}

In the table below we see that there is also no interesting quantum behaviour in the bipartite EPR-steering qualifications, with all values being well above those needed to demonstrate EPR-steering.

\begin{center}
\begin{tabular}{ | c || c | c | c | c | c | c |}
\hline
 \multicolumn{7}{| c |}{Bipartite EPR-steering, loss at 2} \\
 \hline
   &  $EPR_{12}$ & $EPR_{21}$ & $EPR_{23}$ & $EPR_{32}$ & $EPR_{13}$ & $EPR_{31}$  \\ 
 \hline
\hline
 $\chi = 10^{-2}$ & 33.3@127$^o$  & 1.4@78$^o$  & 1.8@80$^{0}$  & 18.3@58$^o$  & 6.7@124$^o$  & 4.9@58$^o$ \\
 \hline
 \hline
\end{tabular}
\end{center}

When we examine tripartite entanglement and inseparability, we also find no evidence of violation of any of the inequalities.

\begin{center}
\begin{tabular}{ | c || c | c | c | c | c | c |}
\hline
 \multicolumn{7}{| c |}{Tripartite entanglement, loss at 2} \\
 \hline
   &  $V_{12}$ & $V_{13}$ & $V_{23}$ & $V_{123}$ & $V_{231}$ & $V_{312}$   \\ 
 \hline
\hline
 $\chi = 10^{-2}$ &14.5@118$^o$ & 40.0@98$^o$   & 13.2@62$^o$ & 30.8@102$^o$  & 8.4@95$^o$ & 33.0@87$^o$ \\
 \hline
 \hline
\end{tabular}
\end{center}

The inequalities for tripartite EPR-steering are also not violated, which shows that the interesting quantum behaviour in this configuration is in the mean-fields, and not in the correlations between modes. This configuration is therefore not a good candidate for the preparation of quantum correlated states of atoms.

\begin{center}
\begin{tabular}{ | c || c | c | c |}
\hline
 \multicolumn{4}{| c |}{Tripartite EPR-steering, loss at 2} \\
 \hline
   &   $OBR_{123}$ & $OBR_{231}$ & $OBR_{312}$  \\ 
 \hline
\hline
 $\chi = 10^{-2}$ & 10.4@124$^o$  & 1.7@84$^o$ & 5.9@60$^o$  \\
 \hline
 \hline
\end{tabular}
\end{center}

\section{Both pumping and damping at the first well}
\label{sec:BH3b1g1line}

In this case the classical non-interacting steady-state solutions are found as
\begin{eqnarray}
\alpha_{1} &=& \frac{\epsilon}{\gamma}, \nonumber \\
\alpha_{2} &=& 0, \nonumber \\
\alpha_{3} &=& -\frac{\epsilon}{\gamma}.
\label{eq:classicp1g1}
\end{eqnarray}
The solutions for $\alpha_{1}$ and $\alpha_{3}$ are of opposite sign in this case, so that the population in the middle well does not increase as a result of destructive interferences, as also seen in a proposal for a four-well atomtronic phase gate~\cite{atomtronic4}. When we simulate the truncated Wigner equations, we find that a steady-state population of $N_{2}\approx 2.9$ for $\chi=0$, which is possible because the atomic modes do not have a sharply defined phase quantum mechanically and therefore the destructive interference is not total.

\begin{figure}[tbhp]
\centering
\begin{minipage}{.5\textwidth}
  \centering
  \includegraphics[width=.9\linewidth,height=0.65\linewidth]{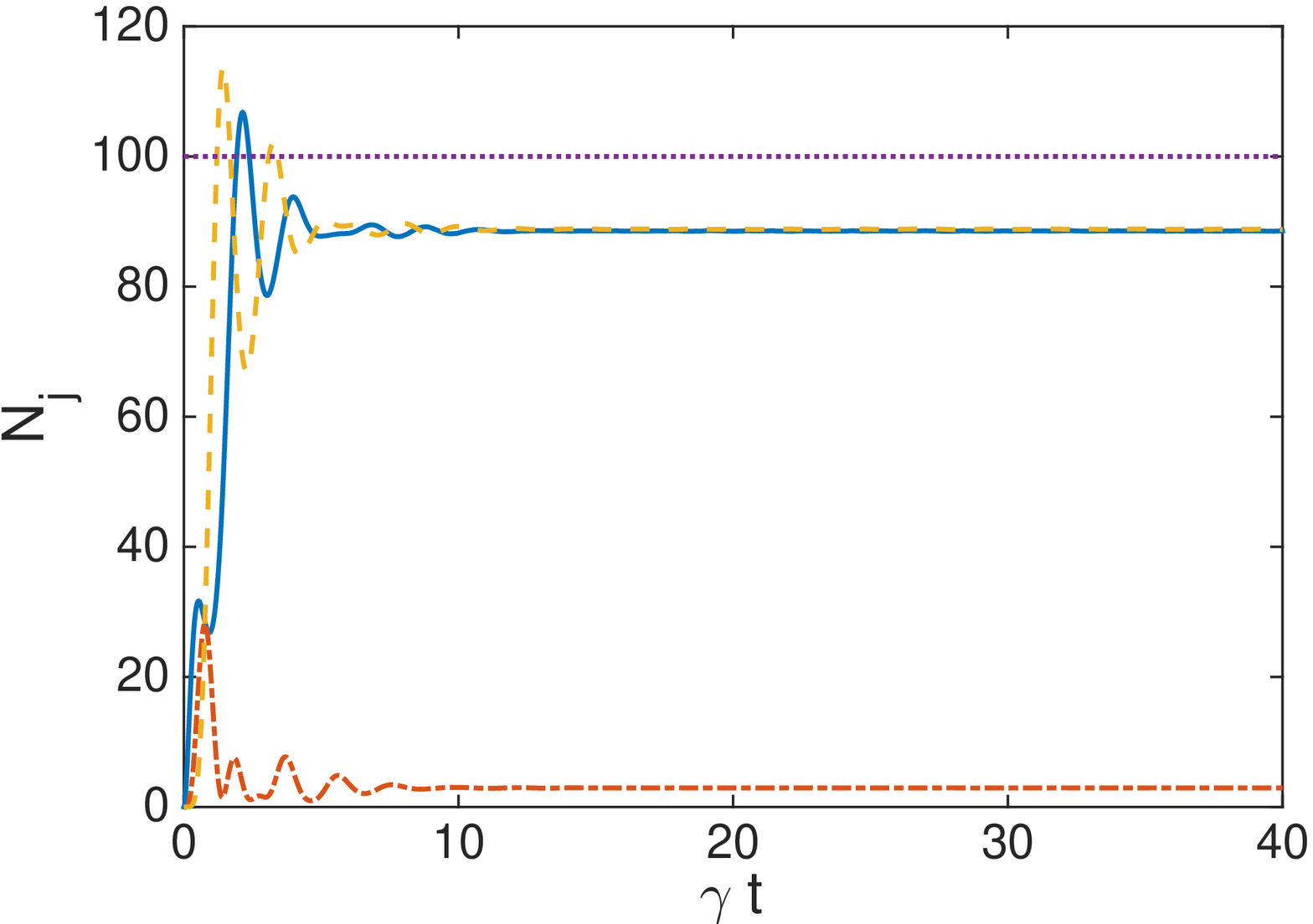}
\end{minipage}%
\begin{minipage}{.5\textwidth}
  \centering
  \includegraphics[width=.9\linewidth,height=0.65\linewidth]{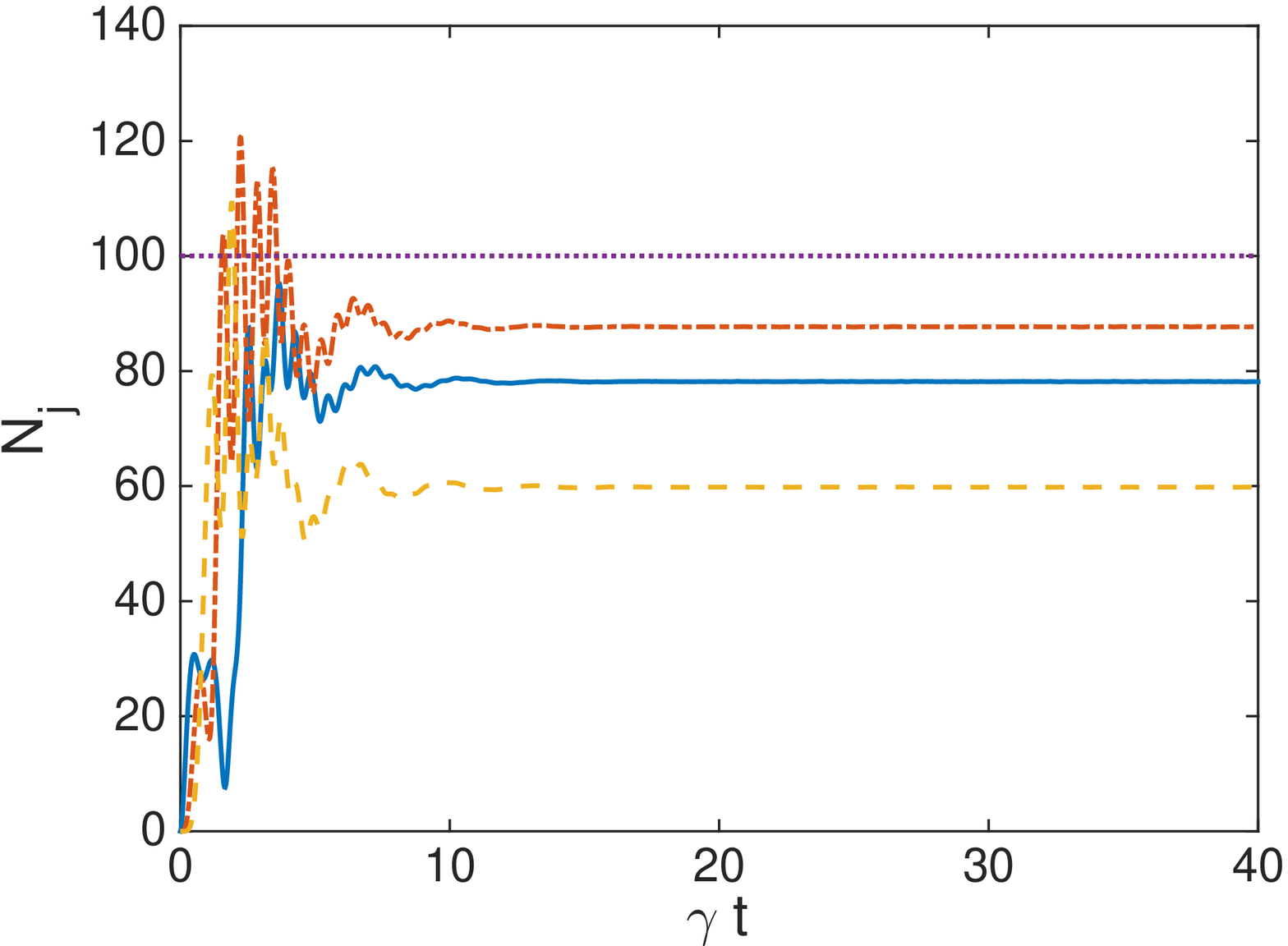}
 \end{minipage}
 \caption{(Color online)(a) The populations in each well as a function of time for $\chi=10^{-3}$ and damping at well 1. The solid line is $N_{1}$, the dash-dotted line is $N_{2}$, and the dashed line is $N_{3}$. The dotted line represents the non-interacting steady-state analytical solutions for modes 1 and 3. The quantities plotted in this and subsequent plots are dimensionless. \newline
 (b) The populations in each well as a function of time for $\chi=10^{-2}$ and damping at well 1. The line styles are the same as in (a).}
 \label{fig:popsg1line}
\end{figure}

As $\chi$ is increased, the population of the middle well increases as shown in Fig.~\ref{fig:popsg1line}, with the other two populations decreasing relative to their non-interacting solutions. The increase in the middle well occupation is due to a change in the mean phase difference between it and the end wells, which is as expected for the Kerr type of nonlinearity, which rotates the phases at different rates depending on the number in each well, and also causes phase diffusion in each well. This effect is seen markedly for $\chi=10^{-2}$, for which the population in the middle well is actually higher than that in either of the two end wells.

The next table shows that this configuration is the optimal of the three for the manufacture of states exhibiting bipartite inseparability and quadrature squeezing. All three quadratures exhibit squeezing in the steady-state for both values of nonlinearity, while only $DS_{12}$ fails to violate the Duan-Simon inequality, and then only for $\chi=10^{-3}$. As seen previously, the correlations tend to improve for the increased $\chi$, although $V(\hat{X}_{3})$ and $DS_{13}$ provide counterexamples.

\begin{center}
\begin{tabular}{ | c || c | c | c | c | c | c |}
\hline
 \multicolumn{7}{| c |}{Bipartite entanglement, loss at 1} \\
 \hline
   &  $V(\hat{X}_{1})$ & $V(\hat{X}_{2})$ & $V(\hat{X}_{3})$ & $DS_{12}$ & $DS_{13}$ & $DS_{23}$  \\ 
 \hline
 \hline
$\chi = 10^{-3}$ & 0.8@2$^o$  & 0.7@93$^o$    & 0.66@7$^o$ & 4.2@95$^o$ & 2.9@5$^o$  & 4.5@44$^o$   \\
\hline
 $\chi = 10^{-2}$ & 0.68@164$^o$ & 0.66@155$^o$    & 0.76@153$^o$   & 2.8@158$^o$  & 3.0@156$^o$  &2.8@153$^o$ \\
 \hline
 \hline
\end{tabular}
\end{center}

\begin{figure}[tbhp]
\includegraphics[width=0.75\columnwidth]{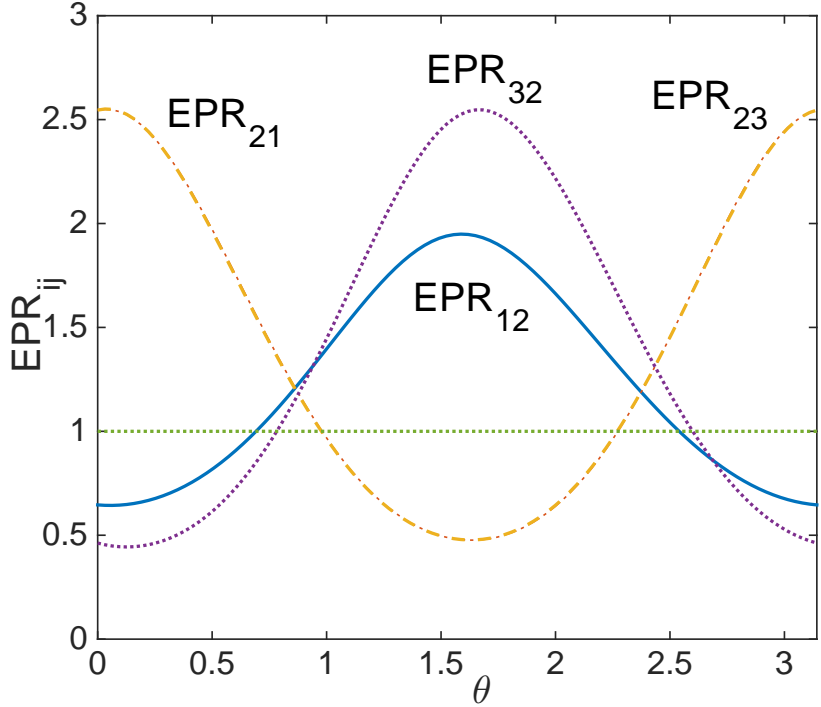}
\caption{(colour online) The steady-state Reid EPR correlations for modes 1 and 2, and 2 and 3,  as a function of quadrature angle, for damping at well 1 and $\chi=10^{-3}$. The pairs $EPR_{12}$, $EPR_{21}$, and $EPR_{23}$, $EPR_{32}$ do not violate the inequality at the same quadrature angles, so that these bipartitions exhibit an asymmetric steering which can be tuned by the angle chosen for measurement. This is not the case for modes 1 and 3, nor for any bipartition for $\chi=10^{-2}$.}
\label{fig:EPR12g1ki3}
\end{figure}

The bipartite EPR-steering results are given in the next table, from which we see that all bipartitions exhibit EPR-steering for both values of the collisional nonlinearity. We do find quadrature dependent asymmetric steering for two of the bipartitions when $\chi=10^{-3}$, as shown in Fig.~\ref{fig:EPR12g1ki3}. This is not the case for $\chi=10^{-2}$, where only symmetric steering is present. We note here that this is the only one of the three configurations considered that shows EPR-steering for all bipartitions, and for both values of $\chi$. Also of interest is the fact that $EPR_{21}$ and $EPR_{23}$ give equal values and that this holds over all quadrature angles, as seen in Fig.~\ref{fig:EPR12g1ki3} for $\chi=10^{-3}$.

\begin{center}
\begin{tabular}{ | c || c | c | c | c | c | c |}
\hline
 \multicolumn{7}{| c |}{Bipartite EPR-steering, loss at 1} \\
 \hline
   &  $EPR_{12}$ & $EPR_{21}$ & $EPR_{23}$ & $EPR_{32}$ & $EPR_{13}$ & $EPR_{31}$  \\ 
 \hline
 \hline
$\chi = 10^{-3}$ & 0.65@4$^o$   & 0.48@93$^o$    & 0.48@93$^o$  & 0.44@7$^o$   & 0.65@2$^o$  & 0.44@7$^o$ \\
\hline
 $\chi = 10^{-2}$ &0.46@164$^o$    & 0.43@155$^o$   & 0.43@155$^o$  & 0.57@153$^o$ & 0.47@164$^o$  & 0.58@153$^o$  \\
 \hline
 \hline
\end{tabular}
\end{center}

When we look at the tripartite entanglement/inseparability correlations, shown in the following table, we see that inseparability is not shown by the $V_{ij}$ values for $\chi=10^{-3}$, with only one of these being less than 4. However, tripartite inseparability is demonstrated by $V_{123}$ and $V_{312}$, although the violations of the inequality are not sufficient to demonstrate genuine tripartite entanglement. For the higher $\chi$ value, inseparability but not entanglement is demonstrated by all six of the correlation functions.

\begin{center}
\begin{tabular}{ | c || c | c | c | c | c | c |}
\hline
 \multicolumn{7}{| c |}{Tripartite entanglement, loss at 1} \\
 \hline
   &  $V_{12}$ & $V_{13}$ & $V_{23}$ & $V_{123}$ & $V_{231}$ & $V_{312}$   \\ 
 \hline
 \hline
$\chi = 10^{-3}$ & 4.2@95$^o$ & 2.9@5$^o$  & 4.4@42$^o$  & 3.7@18$^o$  &  4.4@85$^o$  & 3.8@175$^o$  \\
\hline
 $\chi = 10^{-2}$ & 2.8@156$^o$  & 2.8@156$^o$   & 3.0@153$^o$  & 2.8@39$^o$  & 2.8@156$^o$  & 2.8@156$^o$  \\
 \hline
 \hline
\end{tabular}
\end{center}

When we look at the possibility of two of the three possible wells combining to steer the other, we see in the table below that all three possible pairs can do this for both values of the nonlinearity. All the correlations show values below one at some quadrature angle. Due to the fact that steering is a subset of entanglement, this is sufficient to prove that genuine tripartite entanglement exists in this system. Overall, it is this configuration with pumping and damping at the same well that gives the best performance in terms of quantum correlations.

\begin{center}
\begin{tabular}{ | c || c | c | c |}
\hline
 \multicolumn{4}{| c |}{Tripartite EPR-steering, loss at 1} \\
 \hline
   &   $OBR_{123}$ & $OBR_{231}$ & $OBR_{312}$  \\ 
 \hline
 \hline
$\chi = 10^{-3}$ & 0.65@4$^o$  & 0.48@93$^o$  & 0.44@7$^o$    \\
\hline
 $\chi = 10^{-2}$ & 0.46@164$^o$ & 0.43@155$^o$  & 0.57@153$^o$  \\
 \hline
 \hline
\end{tabular}
\end{center}

\section{Conclusions}
\label{sec:conclusions}

Basing ourselves on recent experimental advances in the manipulation of lattice potentials and trapped atoms, we have proposed and analysed three different configurations of an open pumped and damped Bose-Hubbard trimer. With pumping always at one of the end wells, we find that both the behaviour of the mean fields and the quantum correlations have a qualitative and quantitative dependence on the choice of damped well.
The use of atomic potentials rather than optical cavities gives a freedom in choosing which modes are to be damped that is not available for photonic systems.

We found that the configuration with pumping and damping at opposite ends of the chain is that for which the classical mean-field equations give results which are closest to the quantum solutions of the truncated Wigner approximation. This configuration also exhibits some interesting quantum behaviour, notably in terms of a quadrature angle dependent asymmetric EPR-steering between two pairs of modes.  

The configuration with damping of the middle well is interesting in that the classical equations fail to predict the populations, with both quantitative and qualitative differences with the quantum solutions. Apart from this feature, it does not evidence any interesting behaviour in terms of quantum correlations.

The configuration which gives the best performance in terms of quantum correlations is that with pumping and damping both at the end well. This configuration also exhibits a marked dependence of the middle well population on the collisional nonlinearity. For $\chi=10^{-3}$ this population is heavily suppressed compared to that of the other two modes, whereas it becomes greater than either for $\chi=10^{-2}$. We have explained this feature in terms of phase diffusion. This configuration is the best choice for the manufacture of quantum correlated atomic states.

\section*{Acknowledgments}

This research was supported by the Australian Research Council under the Future Fellowships Program (Grant ID: FT100100515).

\end{document}